\documentclass[twoside,british,1p,11pt]{elsarticle}
\usepackage[T1]{fontenc}
\usepackage[utf8]{inputenc}
\usepackage[table,xcdraw]{xcolor}
\usepackage{pagecolor}
\usepackage{pdfcolmk}
\usepackage{units}
\usepackage{textcomp}
\usepackage{bbding}
\usepackage{threeparttable}
\usepackage{amsmath}
\usepackage{titlesec}
\usepackage{amssymb}
\usepackage{amsthm}
\usepackage{algorithm}
\usepackage{algorithmic}
\usepackage{graphicx}
\usepackage{tikz}                       
\usetikzlibrary{shapes,arrows,positioning}
\usepackage{setspace}
\usepackage{esvect}
\PassOptionsToPackage{normalem}{ulem}
\usepackage{ulem}
\usepackage{blindtext}
\usepackage{multicol}
\usepackage{booktabs}
\usepackage{tabularx}
\usepackage{longtable}
\usepackage{pbox}
\usepackage{csquotes}

\usepackage{caption}
\usepackage{subcaption}

\usepackage{lscape}
\usepackage{hyperref}
\usepackage{makecell}
\usepackage{multirow}

\usepackage{enumitem}
\usepackage{dcolumn}   
\usepackage{bm}        
\usepackage{babel}

\hypersetup{
    colorlinks=true,
    linkcolor=blue,
    filecolor=magenta,      
    urlcolor=cyan,
    pdfborderstyle={/S/U}
}

\theoremstyle{definition}


\DeclareGraphicsExtensions{.pdf,.jpeg,.png,.eps}

\titlespacing*{\section}{0pt}{8pt plus 2pt minus 2pt}{4pt plus 1pt minus 1pt}
\titlespacing*{\subsection}{0pt}{6pt plus 2pt minus 2pt}{2pt plus 1pt minus 1pt}
\titlespacing*{\subsubsection}{0pt}{6pt plus 2pt minus 2pt}{1pt}

\begin{document}

\begin{frontmatter}

\title{Detecting Deceptive Recruitment: A Signal-theoretic Machine Learning Framework for Early Identification of Labour Exploitation}

\author[aff3,aff4]{Sajid Siraj\corref{fn1}}
\ead{s.siraj@leeds.ac.uk}
\author[aff1]{Mahnaz Hosseinzadeh}
\ead{m.hosseinzadeh@sheffield.ac.uk}
\author[aff1]{Amin Vafadarnikjoo}
\ead{a.vafadarnikjoo@sheffield.ac.uk}
\author[aff2]{Shuyang Li}
\ead{s.li.10@bham.ac.uk}

\cortext[fn1]{Corresponding author.}

\address[aff3]{Leeds University Business School, University of Leeds, Leeds, UK}
\address[aff4]{COMSATS University, Wah Campus, Islamabad, Pakistan}
\address[aff1]{Sheffield University Management School, University of Sheffield, Sheffield, UK}
\address[aff2]{Birmingham Business School, University of Birmingham, Birmingham, UK}

\begin{abstract}
Deceptive online job advertisements have emerged as a primary pathway into forced labour, yet systematic detection methods remain underdeveloped due to data scarcity and absence of empirically validated indicators. We formalise this detection challenge as a classification problem under signalling theory, where exploiters transmit costless signals mimicking legitimate communications across textual, visual, and structural dimensions. Using 464 verified cases (164 deceptive, 300 legitimate) collected through anti-slavery charities across nine origin countries and 21 industries, we develop multimodal detection models combining computer vision, natural language processing, and semantic embeddings. Through systematic feature ablation experiments and repeated stratified cross-validation, we demonstrate that individual modalities achieve substantial discriminatory power (ROC-AUC: 0.87--0.97), whilst their integration yields modest further gains. SHAP-based analysis reveals that text quality and domain-specific risk language are the primary discriminators, with readability indices, risk keyword density, and visa sponsorship mentions ranking highest, followed by visual colour and texture features. These production quality gaps reflect resource constraints that prevent exploiters from maintaining professional standards across all communication channels simultaneously. We operationalise findings through a proof-of-concept decision support system providing interpretable risk scores for practitioners. This work demonstrates how rigorous analytical frameworks can address complex humanitarian operations challenges characterised by information asymmetry and limited ground-truth data.
\end{abstract}
\begin{keyword}
Machine learning classification; Labour supply chain exploitation; Decision support systems; Deceptive recruitment detection; Humanitarian operations
\end{keyword}

\end{frontmatter}{}

\section{Introduction}
\label{sec:introduction}

Labour exploitation through deceptive recruitment is a major operational challenge for global supply chains, with an estimated 28 million people in forced labour worldwide \cite{ilo2022,hosseinzadeh2025assessing}. Unlike product risks traceable through established logistics frameworks, it arises largely from information asymmetries in recruitment markets, where vulnerable workers cannot verify employment claims before accepting positions \cite{lebaron2021role,marques2025impact}. This poses a classification problem: distinguishing legitimate from deceptive offers at initial contact, before exploitation occurs \cite{moyo2025investigating}. Audit-based operational risk tools are reactive, intervening only after workers have entered employment \cite{benstead2021detecting}. The Operational Research (OR) community has developed sophisticated methods for supply chain risk detection \cite{yin2023covid,lin2025generative}, quality control \cite{perdikis2024distribution, song2025automatic}, and fraud identification \cite{hoppner2022instance,ma2025can}, yet application to labour recruitment remains limited. We address this gap by formulating deceptive recruitment detection as a supervised classification problem, grounded in information economics and operationalised through machine learning.

The problem suits OR approaches: it involves decision-making under uncertainty with high-dimensional, multimodal data (text, images, metadata), with risk minimisation through early intervention as the operational objective. Unlike typical classification problems, however, forced labour detection operates under severe data scarcity, since verified cases require collaboration with anti-trafficking organisations and victim consent \cite{ramchandani2025unmasking}. Interventions also remain downstream: workplace audits follow exploitation \cite{benstead2021detecting,meehan2021modern}, and digital tools, while improving supply chain visibility, mostly monitor working conditions rather than detecting deceptive signals at source \cite{christ2021blockchain,yu2025modern}. The recruitment stage, where workers first encounter misleading offers, thus remains largely unexamined, creating a need for accessible, theory-driven tools that identify deceptive cues before exploitation occurs.

This study develops a proof-of-concept AI-driven decision support system (DSS) identifying deceptive job advertisements linked to forced labour. Using 464 verified advertisements (164 linked to confirmed forced labour cases) collected through partnerships with anti-slavery organisations in the UK, Nepal, and Bangladesh, we combine machine learning classification with SHAP (SHapley Additive exPlanations) explainability to identify reliable textual, visual, and structural indicators of deception. We formalise detection as a classification problem under signalling theory, in which exploiters transmit costless signals mimicking legitimate communications across multiple dimensions. The DSS returns interpretable risk scores in real time for supply chain auditors, digital platforms, and labour market regulators.

The research addresses three operational questions:

\begin{enumerate}
    \item What observable features in job advertisements (textual, visual, structural) discriminate deceptive from legitimate recruitment offers?
    \item How do standard supervised learning algorithms (Logistic Regression, Random Forest, XGBoost) perform on this classification task when trained on verified ground-truth data, and what performance trade-offs emerge under class imbalance?
    \item How does signalling theory provide a coherent information-theoretic framework for interpreting model outputs and explaining why specific features predict exploitation risk?
\end{enumerate}

This work contributes to OR in humanitarian operations by showing how information-theoretic models can guide feature engineering in resource-constrained classification. SMOTE (Synthetic Minority Oversampling Technique)-based class balancing with nested cross-validation offers a template for data scarcity in sensitive social domains where sample expansion faces ethical constraints. 
That the interpretable NLP+Visual set matches the full multimodal set, with visual features adding only a modest gain over text alone, is consistent with deceptive signals manifesting partly through cross-modal inconsistency: professional text paired with compromised imagery, or the reverse. Practically, the DSS gives auditors, platforms, and regulators a scalable screening tool, shifting intervention from reactive workplace monitoring to proactive communication analysis.

Section~\ref{sec:Related Work} reviews the literature on labour supply chains, digital detection tools, and signalling theory, identifying the research gaps this study addresses; Section~\ref{sec:methodology} presents the methodology.

\section{Literature review}
\label{sec:Related Work}
\subsection{Labour supply chain and deceptive recruitment}
Forced labour in supply chains is increasingly understood as a structural outcome of global sourcing regimes rather than isolated non-compliance: cost minimisation and compressed delivery timelines systematically generate vulnerabilities that enable exploitation \cite{crane2013modern,gold2015modern,lebaron2021role,meehan2021modern}. This literature expanded following regulatory interventions such as the UK Modern Slavery Act \cite{hosseinzadeh2025assessing}, but focuses predominantly on downstream detection and remediation within product and service supply chains \cite{benstead2021detecting,geng2022addressing,bodendorf2023indicators}, leaving the labour supply chain \cite{snyder2026labor}, particularly recruitment, comparatively under-theorised \cite{soundararajan2021humanizing,marques2025impact,moyo2025investigating}. This matters because recruitment is the primary entry point through which structural pressures translate into coercive labour relations.

Deceptive recruitment is now recognised as a central pathway into forced labour: workers are misled about wages, contracts, or conditions and subsequently trapped through debt, document confiscation, or coercion \cite{crane2013modern,crane2022confronting,shepherd2022organizing,simpson2021role}, with intermediaries exploiting migrant vulnerabilities through contract substitution, fee charging, and misinformation \cite{emberson2022adaptations}. These practices remain largely invisible to audit and compliance systems that prioritise workplace conditions over recruitment processes \cite{soundararajan2021humanizing,marques2025impact}, leaving exploitative brokers unchecked \cite{benstead2021detecting}.

Job advertisements are a primary medium through which such deception is transmitted. As the first point of contact they often misrepresent pay, hours, or benefits, especially across borders where verification is difficult \cite{moyo2025investigating,ramchandani2025unmasking}, allowing intermediaries to exploit workers' aspirations for mobility while obscuring risk \cite{keskin2021cracking}; workers who respond may enter arrangements bearing little resemblance to the original offer \cite{christ2021blockchain}. Yet advertisements remain outside due diligence frameworks, leaving the earliest stage of exploitation unmonitored \cite{jiang2023digital}. Existing responses (targeted audits and worker interviews) illuminate recruitment pathways ex post \cite{benstead2021detecting} but are labour-intensive and workplace-focused, and regulatory initiatives such as zero-fee recruitment do not capture misleading signals embedded in advertisements \cite{lebaron2021role,ihrb2019,vzilinskaite2025migration}. A further constraint is the scarcity of verified ground-truth data, with prior studies relying on expert judgement or manual coding of small samples \cite{konrad2017overcoming,volodko2020spotting,moyo2025investigating}.

\subsection{Digital tools for labour supply chain monitoring and recruitment detection}

Digital technologies are increasingly proposed as complements to audits, which often detect exploitation only after harm has occurred \cite{gold2015modern,benstead2018horizontal,benstead2021detecting,stevenson2018modern}. Blockchain, mobile reporting, and traceability systems aim to create tamper-resistant records and direct communication channels that capture recruitment data audits overlook \cite{christ2021blockchain,jiang2023digital,malakar2025digital,yu2025modern}. Adoption, however, is uneven and context-dependent: many initiatives operate at pilot scale and presuppose digital infrastructure, literacy, and worker access often lacking among vulnerable populations \cite{yu2025modern,malakar2025digital}, since migrants targeted by deceptive recruiters frequently cannot engage with complex systems such as blockchain \cite{fletcher2024recruitment}, and most intervene after recruitment, monitoring employment conditions rather than preventing deception at the point of job search \cite{moyo2025investigating}.

This gap is salient given recruiters' reliance on online platforms, where brokers disseminate misleading offers at scale under weak oversight \cite{ramchandani2025unmasking}, and such postings often carry linguistic and contextual cues associated with exploitation \cite{moyo2025investigating,volodko2020spotting}. AI-based research nonetheless remains fragmented, focusing largely on sex trafficking detection using public social media datasets with non-verified labels \cite{keskin2021cracking,li2023detecting,ramchandani2025unmasking}, so empirical work on forced labour indicators embedded in job advertisements remains limited and conceptually underdeveloped \cite{volodko2020spotting,moyo2025investigating}. Adjacent work illustrates the same gap: machine learning models detected fraudulent job advertisements using expert-labelled data \cite{vidros2017automatic}, whilst unsupervised topic modelling (LDA and LSI) was applied to Facebook advertisements in the Dutch labour market in collaboration with law enforcement \cite{cascavilla2022unsupervised}. Both are methodologically informative but limited: the former targets financial fraud rather than labour exploitation, the latter relies on salary discrepancy as a proxy for forced labour outcomes rather than verified cases, achieving F1 of 0.61.

\subsection{Job advertisement and deceptive cues of labour exploitation}
Recent studies examine how job advertisements signal pathways into forced labour. Drawing on UNODC (United Nations Office on Drugs and Crime) indicators \cite{UNODC2018}, observable cues such as promises of accommodation, absent skill requirements, and excessive hours were operationalised and applied to online postings \cite{volodko2020spotting}, an important step toward operationalisation, but one relying on manual coding and frequency counts rather than empirical validation of predictive power. Individual indicators rarely signify exploitation in isolation, and risk emerges from combinations of cues that manual approaches struggle to capture at scale and that remain vulnerable to subjectivity, leaving existing studies largely descriptive.

Qualitative research reinforces these concerns: digital ethnography shows traffickers deploying informal language, emojis, urgency cues, and trusted-brand references in social media job advertisements \cite{moyo2025investigating}, but the absence of verified ground-truth data constrains generalisability and prevents such cues from being systematically deployed for automated screening. Advancing the field therefore requires theory-driven, data-driven methods capable of testing recruitment signals against confirmed exploitation outcomes, which motivates the framework introduced next.

\subsection{Signalling theory and deceptive signalling}
Signalling theory offers a lens for examining how information is transmitted and interpreted under uncertainty \cite{guo2025enhancing}. Developed in labour economics, it explains how a signaller and a receiver communicate under information asymmetry \cite{spence1973job}: signallers with high-quality but unobservable attributes provide credible, costly signals that lower-quality counterparts cannot easily mimic, allowing receivers to identify quality and allocate resources accordingly \cite{bergh2014signalling,spence2002signaling}. The theory has been applied to explain why firms attract investors \cite{connelly2011signaling,wang2021corporate}, secure customer trust \cite{kumar2022hashtag}, or hire strong candidates \cite{krausert2016hrm}.

An important and under-examined stream concerns \emph{deceptive} signalling, that is, signals that appear costly to receivers but are in fact costless \cite{steigenberger2025deceptive}, where lower-quality signallers mimic high-quality signals and conceal their weaknesses. Generative AI and the expanded reach of digital platforms have intensified the problem, making deceptive signals cheaper to produce and faster to disseminate \cite{steigenberger2025deceptive}. Recruitment markets are fertile ground: workers lack reliable information about job quality and conditions, especially for opportunities advertised online or across borders, whilst recruiters and brokers control much of the information shaping workers' perceptions. Here the signal is the job advertisement itself, conveying the sender's intentions and abilities \cite{musteen2010corporate}, and fraudulent recruiters can mimic legitimate communications (polished language, familiar job titles, references to well-known brands) at little cost. Little is known, however, about how exploiters strategically construct and manipulate recruitment signals in digital labour markets, which requires extending signalling theory to costless, multimodal, strategically deceptive signals operating across textual, visual, and structural dimensions.

\subsection{Research gaps and positioning}

The literature reveals four interconnected gaps that this research addresses.

\textbf{Gap 1: Theoretical framework for deception detection.} Signalling theory has been applied to quality disclosure \cite{bergh2014signalling} and supplier selection \cite{wang2021corporate}, but its application to labour recruitment remains underdeveloped, lacking formal models of how exploiters strategically manipulate multimodal signals (text, visuals, structure) to create pooling equilibria in which deceptive offers become indistinguishable from legitimate ones. This limits understanding of \emph{why} certain features predict exploitation risk.

\textbf{Gap 2: Empirically validated detection indicators.} Prior work \cite{volodko2020spotting,moyo2025investigating} identifies potential deception cues through expert judgement and manual coding, but lacks ground-truth validation against verified forced labour outcomes, statistical evidence of discriminatory power, and quantification of effect sizes. Reliance on synthetic \cite{vidros2017automatic} or self-labelled \cite{keskin2021cracking} data in adjacent domains (employment scams, sex trafficking) further limits transferability to labour exploitation.

\textbf{Gap 3: Multimodal feature analysis.} Detection approaches focus predominantly on textual content \cite{volodko2020spotting,li2023detecting}, overlooking visual and structural signals. No prior study systematically compares the relative importance of textual keywords, visual design attributes (colour, brightness, object context), and semantic embeddings in predicting forced labour risk, leaving practitioners without evidence on which signal dimensions to prioritise in screening.

\textbf{Gap 4: Explainable detection systems.} Machine learning has been applied to trafficking detection \cite{ramchandani2025unmasking}, but the black-box nature of complex algorithms limits deployment in humanitarian contexts requiring stakeholder accountability, offering no interpretable account of \emph{why} an advertisement is classified high-risk and preventing auditors and regulators from validating model reasoning or identifying systematic biases.

We address these gaps by (1) formalising deceptive recruitment as a classification problem under signalling theory, providing information-theoretic foundations for feature engineering; (2) constructing and validating models on 464 verified cases (164 forced labour, 300 legitimate) collected through anti-slavery organisations; (3) systematically comparing textual, visual, and structural features using SHAP-based explainability; and (4) developing a proof-of-concept DSS that operationalises the findings with interpretable risk scores. This positions the work at the intersection of operational research methodology, humanitarian operations, and information economics.

\section{Methodology}
\label{sec:methodology}

This section presents the OR framework for deceptive recruitment detection, structured in six components: problem formalisation, data collection, feature engineering, implementation, experimental design for feature ablation studies, and evaluation.

\subsection{Problem formalisation: Signalling theory framework}

We formalise job advertisements as signals transmitted from employers (senders) to jobseekers (receivers) under information asymmetry. Let $\theta \in \{L,H\}$ denote the unobservable quality type of a posting, where $L$ represents deceptive advertisements leading to forced labour and $H$ represents legitimate employment offers. A signal $\mathbf{s} = (s_{\text{text}}, s_{\text{visual}}, s_{\text{struct}})$ comprises textual content, visual attributes, and structural features extracted through natural language processing and computer vision.

In classical signalling equilibria \cite{spence1973job}, high-quality senders differentiate themselves through costly signals satisfying the single-crossing property: $c(s,H) < c(s,L)$, where $c(\cdot)$ denotes signal production cost. However, digital recruitment markets enable costless mimicry where $c(s_{deceptive}, L) \approx c(s_{deceptive}, H) \approx 0$, producing a pooling equilibrium where low-quality recruiters mimic legitimate communications at minimal expense \cite{steigenberger2025deceptive}. Jobseekers facing signal $\mathbf{s}$ cannot reliably compute the posterior probability $P(\theta = L \mid \mathbf{s})$ without analytical tools.

Our classification framework operationalises the receiver's inference problem by constructing $f: \mathcal{S} \rightarrow [0,1]$ estimating $P(\theta = L \mid \mathbf{s}) = f(\mathbf{x}; \boldsymbol{\beta})$, where $\mathbf{x} \in \mathbb{R}^p$ is a $p$-dimensional feature vector and $\boldsymbol{\beta}$ represents learned parameters. The classifier functions as a signal decoder exploiting distributional differences between deceptive and legitimate signals invisible to individual receivers. The operational objective is improving jobseekers' risk estimation, enabling better decision-making under uncertainty.

The framework yields three predictions that we state \emph{a priori}, before estimating any model, and evaluate against the empirical results in Sections~\ref{sec:classification_performance}--\ref{sec:feature_importance}:
\begin{itemize}
\item \textbf{P1 (cost asymmetry).} Signals whose production is bound to underlying resource constraints, and which therefore remain costly for deceptive senders to fake (vocabulary sophistication, domain-specific operational content, image production quality), should carry greater discriminatory weight than surface cues that are cheap to mimic or post-hoc edit (e.g.\ spelling correctness, superficial formatting).
\item \textbf{P2 (cross-modal complementarity).} Because exploiters face binding resource constraints, deceptive advertisements should exhibit inconsistency across modalities; a multimodal feature set should therefore weakly dominate any single-modality set in discriminatory power, rather than yielding large super-additive gains.
\item \textbf{P3 (partial separability).} Observable multimodal signals should partially separate the latent types $\theta\in\{L,H\}$, so that a decoder trained on them attains substantially better-than-chance discrimination despite the pooling equilibrium faced by individual receivers.
\end{itemize}

\subsection{Data collection and verification protocol}

Data collection for forced labour research faces methodological constraints absent in typical OR classification problems: verified positive cases require access to victim testimonies and anti-trafficking case files, imposing strict ethical protocols and limiting sample size. We addressed these constraints through partnerships with two anti-slavery organisations: Causeway (United Kingdom) and Migration Dristi (Nepal and Bangladesh). Both organisations maintain detailed case records with documented victim outcomes, enabling verification that advertisements genuinely led to forced labour rather than financial scams or other forms of fraud.

Between April 2022 and April 2024, practitioners systematically collected 164 high-risk advertisements meeting three criteria: (1) victims responded to the posting and were recruited through the advertised position; (2) subsequent employment conditions met ILO forced labour indicators (threat of penalty, restriction of movement, debt bondage, or withholding of wages); and (3) case documentation included the original advertisement. This verification protocol ensures ground-truth accuracy.

\textbf{Ground-truth provenance.} The class labels were not produced by a post-hoc annotation exercise carried out by the research team, so conventional inter-rater agreement statistics do not apply. The positive class comprises advertisements that the partner anti-slavery organisations had already verified, through their own casework, as having led to a documented forced-labour outcome; the organisations supplied the de-identified advertisements together with this verified status. To protect victims, no victim identities, testimonies, or case files were shared with the research team, and the organisations' internal verification procedures are operational records we are not in a position to document in detail. Each advertisement therefore carries a single authoritative label, determined by the organisation that handled the case rather than by multiple independent research coders, so an inter-rater reliability coefficient is neither available nor meaningful. The negative class was vetted by the partner organisations' experts for reputational standing and regulatory compliance. We regard this externally determined ground truth as a strength: each positive case is a confirmed real-world outcome rather than a proxy or self-reported label, verified by the practitioners closest to the case.

For negative class representation, we assembled 300 legitimate advertisements from verified employers across similar job domains, vetted by charity experts to ensure reputational standing and regulatory compliance. This sampling strategy maintains distributional similarity between classes (same industries, geographic routes, skill levels) while ensuring outcome differences, critical for identifying discriminative features rather than spurious correlations.

\textbf{Advertisement sources.} The advertisements were supplied by the partner organisations as part of documented cases, rather than collected by the research team from named platforms. They are online recruitment advertisements that the workers concerned had encountered. Systematic platform-of-origin metadata was not retained for each item, so we characterise the corpus by its content rather than by source platform. The information available for each advertisement, and used by the models, is what a jobseeker sees at the point of screening: the advertisement text, the accompanying image(s), and the advertised contact method (illustrated in Figure~\ref{fig:synthetic_ad}). Analysis of source-specific performance would require provenance metadata that this dataset does not carry, which we note as a direction for future data collection.

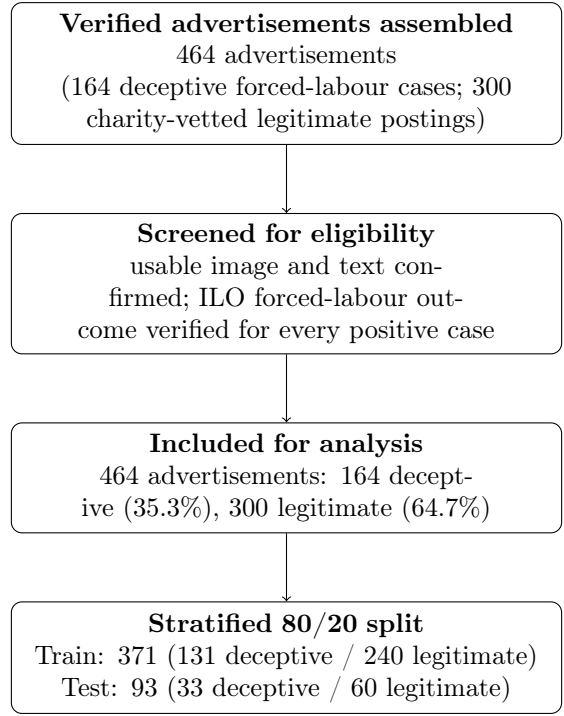
\begin{figure}[!htb]
\centering
\begin{tikzpicture}[node distance=9mm, every node/.style={font=\small},
  box/.style={draw, rounded corners, align=center, text width=7.0cm, inner sep=4pt}]
\node[box] (id) {\textbf{Verified advertisements assembled}\\464 advertisements\\(164 deceptive forced-labour cases; 300 charity-vetted legitimate postings)};
\node[box, below=of id] (scr) {\textbf{Screened for eligibility}\\usable image and text confirmed; ILO forced-labour outcome verified for every positive case};
\node[box, below=of scr] (inc) {\textbf{Included for analysis}\\464 advertisements: 164 deceptive (35.3\%), 300 legitimate (64.7\%)};
\node[box, below=of inc] (spl) {\textbf{Stratified 80/20 split}\\Train: 371 (131 deceptive / 240 legitimate)\\Test: 93 (33 deceptive / 60 legitimate)};
\draw[->] (id) -- (scr);
\draw[->] (scr) -- (inc);
\draw[->] (inc) -- (spl);
\end{tikzpicture}
\caption{Data flow from verified-case assembly to the analysis sample and the stratified train-test split (PRISMA-style).}
\label{fig:data_flow}
\end{figure}

The resulting dataset comprises $n = 464$ advertisements with class distribution $n_L = 164$ (deceptive, 35.3\%) and $n_H = 300$ (legitimate, 64.7\%). Geographic coverage includes nine origin countries and multiple destination regions (Europe, North America, Middle East, East Asia), spanning 21 industry sectors: agriculture, hospitality, care work, automotive, manufacturing, retail, logistics, construction, food production, and information technology. Figure~\ref{fig:data_flow} summarises the flow from assembled cases through screening and inclusion to the stratified train--test split.

\begin{figure}[!htb]
\centering
\fbox{\begin{minipage}{0.86\linewidth}
\small
\textbf{URGENT HIRING --- Factory Helpers Needed ABROAD (No Experience)}\\[2pt]
Good salary guaranteed! \textbf{Free visa + free accommodation} provided. Immediate joining. Limited seats!! Company arranges everything.\\[2pt]
Roles: packing, loading, general helper. Age 20--40.\\[2pt]
Contact on WhatsApp: +xx xxxxx xxxxx --- message now, don't miss this chance!\\[4pt]
\textit{[Accompanying image: low-resolution photograph of a generic warehouse; no company logo, address, or registration detail.]}
\end{minipage}}
\caption{Synthetic, illustrative reconstruction of a deceptive job advertisement, showing the in-situ structure our features target: urgency and scarcity cues, visa/accommodation sponsorship offers, an informal messaging-app contact channel, and low-quality imagery lacking verifiable employer detail. This is a fabricated example for exposition only; no real advertisement, victim, or identifying information is reproduced, in line with safeguarding requirements. Full original advertisements cannot be shared for the same reason.}
\label{fig:synthetic_ad}
\end{figure}

\subsection{Multimodal feature engineering pipeline}

Feature extraction transforms raw advertisement data (images, text, metadata) into feature vectors spanning three modalities. After zero-variance filtering (removing YOLO object-class columns never detected across the 464 advertisements), the retained set comprises 922 dimensions in the full configuration and 154 in the NLP+Visual configuration: 94 visual, 60 engineered textual, and 768 embedding dimensions. The complete feature inventory, with per-category dimension counts, is given in Supplementary Table S4.

\textbf{Visual features} span low-level pixel intensity and colour statistics, mid-level texture and composition metrics (Local Binary Patterns, Grey-Level Co-occurrence Matrix, edge density, sharpness, symmetry), and high-level per-class YOLO object-detection flags. Object detection uses YOLOv8n trained on the COCO 80-class vocabulary; flags represent pixel-pattern matches within that vocabulary rather than semantic concept recognition, a distinction relevant to interpreting individual YOLO-derived features (see Section~\ref{sec:Conclusion}).

\textbf{Textual features} are extracted via multilingual OCR (EasyOCR; Hindi, Urdu, Bangla, English) and cover basic text structure, quality metrics (Flesch, Gunning Fog, Coleman-Liau, and Flesch-Kincaid readability indices, spelling and grammar error rates, sentiment polarity), domain-specific risk keywords (visa sponsorship, seasonal work, accommodation, work permits, informal contact channels such as WhatsApp), psychological triggers, NRC Lexicon emotion intensities, and composite risk indicators.

\textbf{Structural features} comprise BERT-base-multilingual-cased embeddings (768 dimensions) encoding contextual word relationships and rhetorical structures, used in the comprehensive baseline but excluded from the interpretable NLP+Visual configuration due to inference-time computational constraints.

\subsection{Supervised classification models}

We implement three supervised learning algorithms enabling comparative assessment across feature configurations; results tables abbreviate them as LR, RF, and XGB respectively. \textbf{Logistic Regression (LR)} estimates deception probability via maximum likelihood with $L_2$ regularisation ($C \in \{0.01, 0.1, 1, 10\}$), providing maximum interpretability through transparent coefficient inspection, which is critical for operational deployment. \textbf{Random Forest (RF)} constructs bootstrap-aggregated decision tree ensembles with variable ensemble size ($n\_estimators \in \{50, 100, 200\}$) and tree complexity ($max\_depth \in \{\texttt{None}, 10, 20\}$), naturally handling nonlinear relationships though with reduced interpretability. \textbf{XGBoost (XGB)} implements sequential tree construction with tuned boosting rounds ($n\_estimators \in \{50, 100, 200\}$) and learning rate ($learning\_rate \in \{0.01, 0.1, 0.3\}$), often achieving superior performance on imbalanced data. 

To address the 35.3\%/64.7\% class imbalance, we apply SMOTE (Synthetic Minority Oversampling Technique) during training, generating synthetic minority class examples by interpolating between existing instances, applied only to training folds to ensure evaluation reflects true performance on unseen imbalanced distributions.

\subsection{Experimental design: Systematic feature ablation studies}

A central methodological contribution is the systematic evaluation of feature modality contributions through controlled ablation experiments. We implement six distinct experimental configurations isolating and combining feature modalities, enabling direct quantification of each modality's marginal contribution through performance comparison.

\textbf{Configuration 0: Keyword-only baseline} ($p = 22$) uses only the domain-specific risk keyword and psychological trigger binary features, representing the simplest possible theory-driven screening rule. This provides a reference point against which the value of adding visual and quality features can be assessed. \textbf{Configuration 1: Visual features only} ($p = 94$) isolates visual low-level, mid-level, and high-level features, testing whether image attributes alone provide sufficient discriminatory power. \textbf{Configuration 2: NLP features only} ($p = 60$) includes text basic, text quality, risk keywords, psychological triggers, emotions, and composite risk indicators, evaluating textual signal detection independent of visual and semantic information. \textbf{Configuration 3: Embeddings only} ($p = 768$) uses solely BERT-generated semantic representations, assessing whether deep contextual embeddings capture exploitation patterns without explicit feature engineering. \textbf{Configuration 4: NLP + Visual} ($p = 154$) combines textual and visual features while excluding embeddings, testing whether engineered features suffice without deep learning representations. \textbf{Configuration 5: All features} ($p = 922$) serves as the comprehensive baseline, incorporating visual, textual, and structural features simultaneously to establish maximum achievable performance. All feature counts reflect post-filtering values after zero-variance columns are removed.

For each configuration $c \in \{0,\ldots,5\}$, we train all three algorithms independently, yielding 16 distinct model-configuration combinations (three algorithms across five full feature configurations, plus the keyword-only logistic regression baseline). The ablation framework addresses three analytical objectives: (i) \textit{Modality sufficiency}: do visual or textual features alone achieve competitive performance, or is multimodal integration necessary? (ii) \textit{Engineered vs. learned features}: can BERT embeddings replace manual feature engineering, or do explicit indicators provide complementary information? (iii) \textit{Interaction effects}: does combining modalities yield superadditive performance gains, suggesting that deceptive signals manifest through cross-modal inconsistencies?

\subsection{Evaluation framework}

Model evaluation follows a rigorous train-test protocol applied uniformly across all experimental configurations. We implement stratified 80/20 train-test splitting, yielding 371 training instances (131 deceptive, 240 legitimate) and 93 test instances (33 deceptive, 60 legitimate). Stratification ensures test set class proportions match the full dataset (35.3\%/64.7\%), preventing evaluation bias from unrepresentative sampling.

\textbf{SMOTE application.} SMOTE (Synthetic Minority Oversampling Technique) is applied exclusively to the training partition after the stratified split, generating synthetic minority-class examples by interpolating between existing instances. This produces a balanced 480-instance training set whilst preserving the natural 35.3\%/64.7\% imbalance in the held-out test partition, ensuring that reported performance metrics reflect realistic operating conditions rather than an artificially balanced distribution.

\textbf{Hyperparameter tuning.} For each algorithm-configuration pair, we conduct grid search with 5-fold stratified cross-validation on the training set, selecting hyperparameter configurations maximising ROC-AUC score. All continuous features undergo z-score standardisation (zero mean, unit variance) fitted on training data and applied to test data, preventing data leakage.

\textbf{Uncertainty quantification.} We compute 95\% bootstrap confidence intervals (2,000 resampling iterations, percentile method) on both ROC-AUC and PR-AUC for each held-out test evaluation. To provide generalisation estimates less sensitive to a single partition, we additionally conduct repeated stratified $k$-fold cross-validation (5 folds $\times$ 10 repeats, $k \times r = 50$ outer folds) with inner 5-fold GridSearchCV for hyperparameter selection within each outer fold, constituting a true nested cross-validation. SMOTE is re-applied inside each training fold to prevent synthetic samples from appearing in validation partitions. This repeated CV procedure yields mean $\pm$ standard deviation performance estimates across 50 fold realisations.

\begin{figure}[!htb]
\centering
\includegraphics[width=0.85\linewidth]{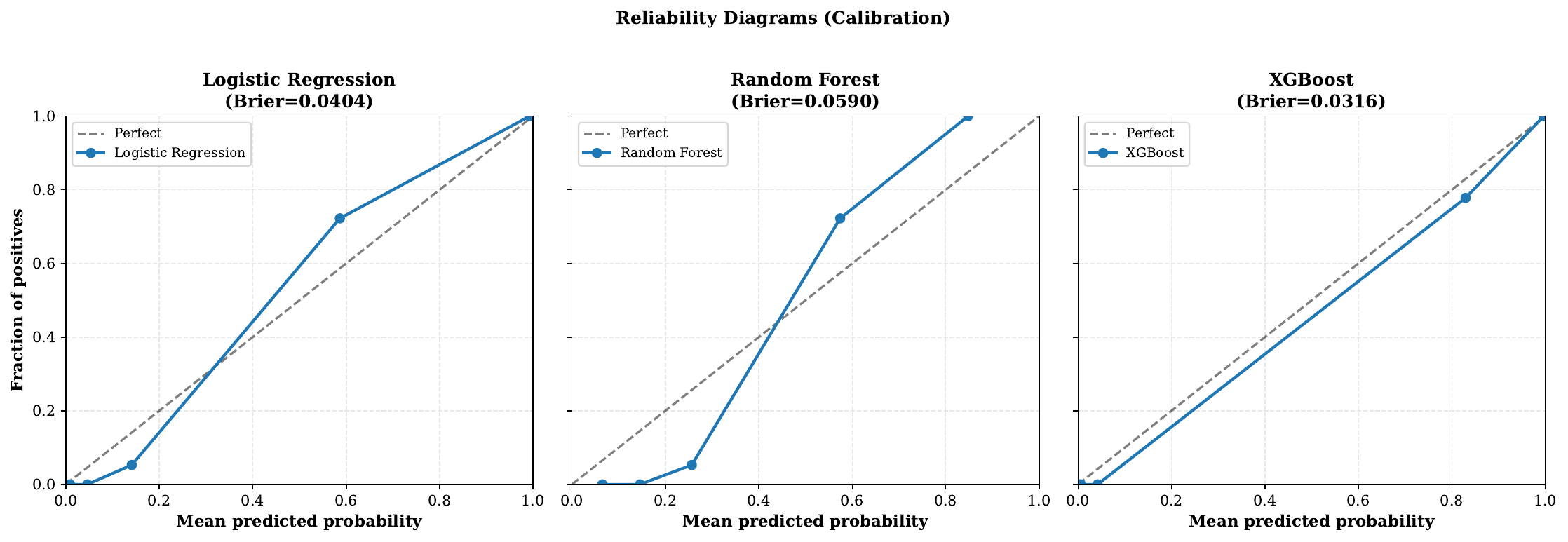}
\caption{Reliability diagrams (five equal-frequency bins) for the three classifiers on the held-out test set (NLP+Visual configuration). Proximity to the diagonal indicates well-calibrated predicted probabilities; the corresponding Brier scores are reported in Table~\ref{tab:repeated_cv}.}
\label{fig:calibration}
\end{figure}

\textbf{Calibration.} We report the Brier score (mean squared error between predicted probabilities and binary outcomes; lower is better) and produce reliability diagrams (five equal-frequency bins) for each model (Figure~\ref{fig:calibration}), enabling assessment of probabilistic output quality beyond discrimination metrics.

\textbf{Performance metrics.} We employ six complementary metrics: Accuracy (overall correctness), Precision (proportion of flagged advertisements truly deceptive), Recall (sensitivity; reflects operational priority of minimising missed exploitation cases), F1-Score (harmonic mean of precision and recall, preferable to accuracy under class imbalance), ROC-AUC (Area Under the Receiver Operating Characteristic curve; primary discrimination metric, threshold-invariant \cite{nayak2021comprehensive}), and PR-AUC (Area Under the Precision-Recall curve; co-primary metric, more informative than ROC-AUC under class imbalance as it explicitly penalises false positives relative to true positives).

\subsection{Explainability analysis: SHAP decomposition}

To ensure model transparency and operational deployability, we implement SHAP analysis \cite{lundberg2020local, lundberg2017unified} on the best-performing model. SHAP assigns each feature a Shapley value representing its marginal contribution to predictions, computed by averaging contributions across all possible feature subsets.

We compute SHAP values using algorithm-specific implementations: \texttt{LinearExplainer} for Logistic Regression and \texttt{TreeExplainer} for Random Forest and XGBoost. Analysis yields: (1) \textbf{Global feature importance}: average absolute SHAP values across all test instances, ranking features by systematic influence on predictions; (2) \textbf{Local explanations}: instance-specific SHAP values showing why individual advertisements receive particular risk scores; (3) \textbf{Directional effects}: sign of SHAP value indicates whether feature presence increases or decreases predicted exploitation risk; (4) \textbf{Category-level contributions}: aggregating SHAP values within feature categories quantifies modality-specific explanatory power, validating ablation study findings through complementary analysis.

\subsection{Statistical comparison framework}

To rigorously compare performance across experimental configurations, we report: (1) \textbf{Within-configuration model comparison}: for each configuration, we identify the best-performing algorithm via ROC-AUC maximisation, reporting full confusion matrices and all five performance metrics; (2) \textbf{Cross-configuration analysis}: we compute performance differences $\Delta_{\text{AUC}} = \text{AUC}_{\text{all}} - \text{AUC}_{\text{subset}}$ between the comprehensive model and ablated variants, quantifying marginal contributions of excluded modalities; (3) \textbf{Relative performance preservation}: we calculate $\rho = \frac{\text{AUC}_{\text{subset}}}{\text{AUC}_{\text{all}}}$ to assess what proportion of maximum performance each subset achieves.

\section{Results}
\label{sec:Results}

Findings are organised in three parts: descriptive feature analysis (Section~\ref{sec:descriptive_analysis}), classification performance across experimental configurations (Section~\ref{sec:classification_performance}), and SHAP-based feature importance revealing discriminatory mechanisms (Section~\ref{sec:feature_importance}).

\subsection{Descriptive analysis: Univariate feature distributions}
\label{sec:descriptive_analysis}

We first establish baseline distributional differences between deceptive ($n=164$) and legitimate ($n=300$) advertisements through univariate tests, confirming that observable features carry discriminatory signal. Full results for continuous features (t-tests with Cohen's $d$) and binary features (chi-square) appear in Supplementary Tables S1 and S2; key findings follow.

\subsubsection{Key distributional patterns}

\textbf{Visual features} show medium effect sizes for texture and sharpness rather than colour. Deceptive advertisements are less sharp (Laplacian variance $d = -0.51$, $p < 0.001$), with lower textural contrast (GLCM $d = -0.42$) and edge density ($d = -0.31$), consistent with low-resolution, heavily compressed, or screenshot graphics; brightness and saturation show no significant differences ($p > 0.05$). Object detection is more nuanced: airplane imagery is more prevalent in deceptive postings (4.9\% vs. 0\%, $p < 0.001$), whilst books appear \textit{more} often in legitimate ones (26.0\% vs. 8.5\%, $p < 0.001$), suggesting richer, contextually appropriate imagery in professionally produced content. Boats, backpacks, and traffic signage showed no significant differences ($p > 0.10$).

\textbf{Textual features} yield the strongest discriminator in the feature set: the Coleman-Liau readability index ($d = -1.21$, $p < 0.001$), with deceptive advertisements scoring substantially lower, indicating simpler, shorter-word vocabulary. Flesch readability ($d = 0.87$) and Flesch-Kincaid Grade ($d = -0.60$) corroborate this. Notably, \textit{spelling error rate does not differ significantly} between classes ($p = 0.107$), challenging the intuitive assumption that deceptive ads contain more typos; grammar error rate ($d = 0.55$) and word count ($d = -0.51$) discriminate more reliably. Risk keywords show the most extreme prevalence gaps: visa/sponsorship terminology (48.8\% vs. 3.7\%, $\chi^2 = 134.0$) and work permit references (28.7\% vs. 0.0\%, $\chi^2 = 92.5$) are near-exclusive to deceptive postings, alongside informal contact channels (WhatsApp: 24.4\% vs. 7.0\%).

\textbf{Emotional content} is counter-intuitive: all positive emotions (anticipation, $d = -0.66$; trust, $d = -0.54$; joy, $d = -0.54$) are significantly \textit{lower} in deceptive advertisements, as legitimate ones use richer affective language to persuade qualified candidates. The pattern extends to psychological triggers: money terms (68.0\% vs. 47.6\%) and urgency language (38.0\% vs. 27.4\%) are \textit{more common} in legitimate advertisements ($p < 0.001$ and $p = 0.029$). Deceptive advertisements are thus affectively impoverished, terse and emotionally neutral, rather than manipulative as originally hypothesised.

These univariate differences are systematic across visual, textual, and emotional dimensions, but differ substantively from prior qualitative accounts. Critically, no single feature separates the classes perfectly, since even the strongest discriminator (Coleman-Liau, $d = -1.21$) overlaps, validating the multivariate approach: reliable detection requires joint consideration of multiple signals rather than thresholds on individual features.

\subsection{Classification performance across experimental configurations}
\label{sec:classification_performance}

This subsection presents classification results across six experimental configurations designed to systematically evaluate the discriminatory power of different feature modalities and establish a keyword-based baseline. Each full configuration was evaluated using three supervised learning algorithms (Logistic Regression, Random Forest, and XGBoost) trained with SMOTE-based class balancing and hyperparameter optimisation via 5-fold cross-validation. Performance is assessed on the held-out test set (93 advertisements: 33 deceptive, 60 legitimate). Bootstrap 95\% confidence intervals for ROC-AUC and PR-AUC are estimated over 2{,}000 resamples; calibration quality is measured via the Brier score (lower is better). Generalisation stability is further evaluated via 5$\times$10 repeated stratified cross-validation on the full training set (Table~\ref{tab:repeated_cv}).

\subsubsection{Keyword-only baseline and comprehensive feature set}

A keyword-only logistic regression baseline on 22 domain-specific risk keyword features (Config.~0) achieves ROC-AUC 0.867 [95\% CI: 0.776--0.940] and PR-AUC 0.806. Simple keyword matching therefore carries substantive discriminatory power, yet the full multimodal models add a consistent $+$0.13 ROC-AUC, justifying the further feature engineering.

The comprehensive model using all 922 features (visual 94, engineered textual 60, and BERT embeddings 768) establishes the discriminatory ceiling. Table~\ref{tab:model_performance_all} compares the three algorithms and Table~\ref{tab:confusion_all} gives the corresponding confusion matrices.
\textbf{XGBoost discriminated best}, with ROC-AUC 0.994 [95\% CI: 0.983--1.000], PR-AUC 0.990, and the lowest Brier score overall (0.034); Logistic Regression attained the highest precision (1.000) and F1 (0.935), with Random Forest matching its recall (0.879). On the held-out test set XGBoost classified 28 of 33 deceptive advertisements correctly (five false negatives) and 59 of 60 legitimate ones (one false positive); all derived metrics in Tables~\ref{tab:model_performance_all} and~\ref{tab:confusion_all} follow from these counts.

\begin{table}[htbp]
\centering
\caption{Model performance comparison for comprehensive feature set (all $p=922$ features)}
\label{tab:model_performance_all}
\small
\begin{tabular}{lccccccc}
\toprule
\textbf{Model} & \textbf{Acc.} & \textbf{Prec.} & \textbf{Rec.} & \textbf{F1} & \textbf{ROC-AUC [95\% CI]} & \textbf{PR-AUC [95\% CI]} & \textbf{Brier} \\
\midrule
LR & \textbf{0.957} & \textbf{1.000} & \textbf{0.879} & \textbf{0.935} & 0.988 [0.968--1.000] & 0.982 [0.952--1.000] & 0.036 \\
RF       & 0.946 & 0.967 & \textbf{0.879} & 0.921 & 0.993 [0.977--1.000] & 0.989 [0.963--1.000] & 0.096 \\
XGB             & 0.935 & 0.966 & 0.848 & 0.903 & \textbf{0.994 [0.983--1.000]} & \textbf{0.990 [0.969--1.000]} & \textbf{0.034} \\
\bottomrule
\end{tabular}
\begin{tablenotes}
\small
\item \textit{Note:} Metrics on held-out test set ($n=93$: 33 deceptive, 60 legitimate). 95\% CIs via percentile bootstrap (2{,}000 iterations). All models trained with SMOTE (training folds only) and 5-fold GridSearchCV. Bold = best value per metric. LR = Logistic Regression, RF = Random Forest, XGB = XGBoost.
\end{tablenotes}
\end{table}

\begin{table}[htbp]
\centering
\caption{Confusion matrices on the held-out test set ($n=93$: 33 deceptive, 60 legitimate) for the comprehensive feature set ($p=922$). All metrics in Table~\ref{tab:model_performance_all} are derived from these counts.}
\label{tab:confusion_all}
\small
\begin{tabular}{lcccc}
\toprule
\textbf{Model} & \textbf{TP} & \textbf{FN} & \textbf{FP} & \textbf{TN} \\
\midrule
LR & 29 & 4 & 0 & 60 \\
RF       & 29 & 4 & 1 & 59 \\
XGB             & 28 & 5 & 1 & 59 \\
\bottomrule
\end{tabular}
\begin{tablenotes}
\small
\item \textit{Note:} Positive class = deceptive. TP$+$FN $=33$ (deceptive) and FP$+$TN $=60$ (legitimate) for every model, confirming arithmetic consistency. LR = Logistic Regression, RF = Random Forest, XGB = XGBoost.
\end{tablenotes}
\end{table}

\begin{table}[htbp]
\centering
\caption{Generalisation under 5$\times$10 repeated stratified cross-validation (NLP+Visual, $p=154$)}
\label{tab:repeated_cv}
\small
\begin{tabular}{lcccccc}
\toprule
\textbf{Model} & \textbf{Accuracy} & \textbf{F1-Score} & \textbf{ROC-AUC} & \textbf{PR-AUC} & \textbf{Brier} \\
\midrule
LR & $0.915 \pm 0.023$ & $0.880 \pm 0.033$ & $0.968 \pm 0.016$ & $0.959 \pm 0.015$ & $0.062 \pm 0.015$ \\
RF       & $0.945 \pm 0.022$ & $0.920 \pm 0.033$ & $0.982 \pm 0.010$ & $0.974 \pm 0.013$ & $0.065 \pm 0.008$ \\
XGB             & $0.941 \pm 0.024$ & $0.916 \pm 0.035$ & $0.982 \pm 0.012$ & $0.972 \pm 0.019$ & $\mathbf{0.048 \pm 0.021}$ \\
\bottomrule
\end{tabular}
\begin{tablenotes}
\small
\item \textit{Note:} Mean $\pm$ std across 50 outer folds (5 repeats $\times$ 10 folds). XGBoost and Random Forest achieve near-identical AUC with XGBoost yielding superior calibration. LR = Logistic Regression, RF = Random Forest, XGB = XGBoost.
\end{tablenotes}
\end{table}

\subsubsection{Individual modality performance and combined configurations}

We trained models on isolated feature subsets to quantify each modality's contribution; Table~\ref{tab:cross_config_comparison} summarises all six configurations.

\textbf{Visual features alone} (94) achieved ROC-AUC 0.866 [CI: 0.782--0.939] with XGBoost, 87\% of maximum performance ($\rho = 0.871$) and closely comparable to the keyword-only baseline (0.868) on a fundamentally different modality, confirming that image quality alone encodes substantial risk signal and enabling lightweight image-only screening. \textbf{Engineered NLP features alone} (61) reached 0.967 [CI: 0.931--0.992] with Logistic Regression, 97\% of maximum ($\rho = 0.972$), substantially outperforming visual-only models with fewer features, and contradicting the initial hypothesis that visual features are primary discriminators.

\textbf{BERT embeddings alone} (768 dimensions) achieved 0.958 [CI: 0.916--0.987] with Random Forest, underperforming the 61-feature engineered NLP set by $\Delta_{\text{AUC}} = -0.009$ despite constituting 83\% of the all-features set by dimension, confirming that domain-specific feature engineering captures the semantic signal at least as well as deep contextual embeddings. \textbf{Combined NLP+Visual} (154 dimensions) achieved 0.992 [CI: 0.978--1.000] with XGBoost, within 0.3 percentage points of the comprehensive all-features model (0.994) while using 16.7\% of its features, and recalling all 33 deceptive advertisements (sensitivity 1.00) at PR-AUC 0.985: the interpretable set attains performance statistically indistinguishable from the embeddings-augmented model, at a fraction of the computational cost and without BERT's opacity.

\begin{table}[htbp]
\centering
\caption{Cross-configuration performance comparison: feature modality contributions}
\label{tab:cross_config_comparison}
\small
\begin{tabular}{lccccccc}
\toprule
\textbf{Configuration} & \textbf{$p$} & \textbf{Best Model} & \textbf{ROC-AUC [95\% CI]} & \textbf{PR-AUC} & \textbf{F1} & \textbf{$\rho$} \\
\midrule
All Features                & 922 & XGB & \textbf{0.994} [0.983--1.000] & \textbf{0.990} & 0.903 & 1.000 \\
NLP+Visual (interpretable) & 154 & XGB & 0.992 [0.978--1.000] & 0.985 & \textbf{0.943} & 0.997 \\
NLP Only                    & 60  & LR & 0.967 [0.931--0.992] & 0.952 & 0.857 & 0.972 \\
Embeddings Only             & 768 & RF & 0.958 [0.916--0.987] & 0.920 & 0.825 & 0.963 \\
Keyword Baseline (Config.~0) & 22 & LR & 0.867 [0.776--0.940] & 0.806 & 0.739 & 0.872 \\
Visual Only                 & 94  & XGB & 0.866 [0.782--0.939] & 0.790 & 0.710 & 0.871 \\
\bottomrule
\end{tabular}
\begin{tablenotes}
\small
\item \textit{Note:} Configurations ranked by ROC-AUC. $\rho$ = relative to the comprehensive all-features configuration. Config.~0 uses Logistic Regression on 23 risk keyword features only. 95\% CIs via bootstrap (2{,}000 iterations). LR = Logistic Regression, RF = Random Forest, XGB = XGBoost.
\end{tablenotes}
\end{table}

\subsubsection{Decision-level evaluation: screening-queue simulation}
ROC-AUC summarises ranking quality but not how screening \emph{decisions} would change in practice. We therefore simulate a screening queue on the held-out test set ($n=93$; 33 deceptive), in which an analyst with a limited review budget inspects advertisements in priority order, comparing three strategies: the multimodal model (XGBoost, NLP+Visual); a keyword rule ranking advertisements by their count of risk-keyword and psychological-trigger flags (the simplest theory-driven heuristic, using the same 23 features as Config.~0); and random ordering.

Table~\ref{tab:screening_queue} reports the outcome. At a review budget equal to test-set prevalence (33 advertisements), the multimodal model surfaces 30 of 33 deceptive advertisements (91\% recall at 91\% precision), against 16 (48\%) for the keyword rule and about 12 (35\%) for random ordering. Fixing a target detection rate widens the gap in analyst effort: recovering 90\% of deceptive cases requires reviewing 32 advertisements with the model but 79 with the keyword rule, a 2.5-fold workload reduction for the same case-finding rate, and the model dominates at every review depth. The performance gap in Section~\ref{sec:classification_performance} therefore translates into materially different operational decisions: at any fixed analyst capacity, the model identifies substantially more exploitation-linked advertisements than the keyword screening characterising current practice.

\begin{table}[htbp]
\centering
\caption{Screening-queue simulation on the held-out test set ($n=93$; 33 deceptive). An analyst reviews advertisements in priority order under each strategy.}
\label{tab:screening_queue}
\small
\begin{tabular}{lccc}
\toprule
 & \textbf{Multimodal model} & \textbf{Keyword rule} & \textbf{Random} \\
\midrule
Deceptive caught in first 33 reviews & 30 (91\%) & 16 (48\%) & $\approx$12 (35\%) \\
Reviews to catch 50\% of cases       & 17 & 36 & 47 \\
Reviews to catch 80\% of cases       & 28 & 72 & 75 \\
Reviews to catch 90\% of cases       & 32 & 79 & 83 \\
Reviews to catch 100\% of cases      & 37 & 89 & 91 \\
\bottomrule
\end{tabular}
\begin{tablenotes}
\small
\item \textit{Note:} Multimodal model $=$ XGBoost (NLP+Visual). Keyword rule ranks advertisements by the count of 23 risk-keyword and psychological-trigger flags. Random $=$ expected value at the test-set prevalence (35.3\% deceptive).
\end{tablenotes}
\end{table}

\subsection{Feature importance and discriminatory mechanisms}
\label{sec:feature_importance}

To identify which signals drive performance, we apply SHAP (SHapley Additive exPlanations) analysis to the XGBoost model from the NLP+Visual configuration (ROC-AUC 0.992), which achieves near-maximum performance whilst remaining fully interpretable. Table~\ref{tab:shap_feature_importance} gives the top 5 features by mean absolute SHAP value; the full top-15 ranking is in Supplementary Table S3.

\begin{table}[htbp]
\centering
\caption{Top 5 features by SHAP importance in NLP+Visual configuration (XGBoost)}
\label{tab:shap_feature_importance}
\small
\begin{tabular}{clp{4.5cm}cc}
\toprule
\textbf{Rank} & \textbf{Feature} & \textbf{Description} & \textbf{Category} & \textbf{Mean SHAP} \\
\midrule
1 & \texttt{coleman\_liau\_index}       & CLI readability score         & Text Quality          & 1.853 \\
2 & \texttt{risk\_keyword\_density}     & Keywords per word             & Composite Risk        & 1.227 \\
3 & \texttt{visa\_sponsorship\_mentioned} & Visa/sponsorship mentioned  & Risk Keywords         & 0.882 \\
4 & \texttt{red\_mean}                  & Mean red channel intensity    & Visual Low-Level      & 0.535 \\
5 & \texttt{unique\_colors}             & Colour diversity (pixel count)& Visual Low-Level      & 0.474 \\
\bottomrule
\end{tabular}
\begin{tablenotes}
\small
\item \textit{Note:} Features ranked by mean absolute SHAP values (scale reflects XGBoost raw outputs). All features are directly interpretable. The top three features are textual; visual features first appear at rank 4.
\end{tablenotes}
\end{table}

\subsubsection{Category-level contributions}

Table~\ref{tab:shap_category_contributions} aggregates SHAP contributions by feature category. \textbf{Textual features dominate with 63.9\% of explanatory power}, led by Text Quality (22.3\%), Basic Text (14.1\%), Composite Risk Indicators (12.4\%), and Domain-Specific Risk Keywords (9.1\%); visual features account for the remaining 36.0\% (Low-Level 18.9\%, Mid-Level 12.1\%, High-Level 5.0\%). This reversal of the expected visual dominance, with NLP-only models reaching 97\% of maximum performance against visual-only models' 87.1\%, establishes textual signal as the primary discriminator.

Within text, quality metrics (Coleman-Liau, grammar error rate, Flesch readability) outweigh domain-specific keywords (22.3\% vs. 9.1\%), confirming that \textit{how} advertisements are written discriminates better than \textit{what} terms they contain; composite risk indicators (12.4\%) form a secondary signal. Visual contribution comes through low-level texture and colour diversity (18.9\%) rather than high-level object detection (5.0\%), consistent with the univariate finding that YOLO classes showed limited significance.

\begin{table}[htbp]
\centering
\caption{Feature category contributions to model explanatory power (NLP+Visual configuration)}
\label{tab:shap_category_contributions}
\small
\begin{tabular}{lcc}
\toprule
\textbf{Category} & \textbf{Contribution (\%)} & \textbf{Modality} \\
\midrule
Text Quality Features          & 22.3\% & NLP \\
Low-Level Visual Features      & 18.9\% & Visual \\
Basic Text Features            & 14.1\% & NLP \\
Composite Risk Indicators      & 12.4\% & NLP \\
Mid-Level Visual Features      & 12.1\% & Visual \\
Domain-Specific Risk Keywords  & 9.1\%  & NLP \\
Emotion Features               & 5.3\%  & NLP \\
High-Level Visual Features     & 5.0\%  & Visual \\
Psychological Triggers         & 0.7\%  & NLP \\
\midrule
\textbf{NLP Total}             & \textbf{63.9\%} & \\
\textbf{Visual Total}          & \textbf{36.0\%} & \\
\bottomrule
\end{tabular}
\begin{tablenotes}
\small
\item \textit{Note:} Percentages represent relative contribution measured via mean absolute SHAP values (XGBoost, NLP+Visual configuration). NLP features dominate despite visual features appearing at ranks 4--5 in the individual feature ranking, because the NLP category contains more high-contributing features overall.
\end{tablenotes}
\end{table}

\subsubsection{Discriminatory mechanisms and signalling interpretation}

SHAP analysis reveals three mechanisms by which deceptive advertisements reveal themselves, each interpretable through information economics.

\textbf{Readability and vocabulary as primary competence signals.} The Coleman-Liau Index (mean SHAP 1.853) leads by a substantial margin, followed by composite risk keyword density (1.227) and visa/sponsorship mention (0.882). Measuring word and sentence length rather than syllabic complexity, Coleman-Liau captures vocabulary sophistication: legitimate advertisements use the longer, more complex words of professional register, deceptive ones shorter and simpler vocabulary. This is an involuntary competence signal: exploiters cannot systematically elevate vocabulary without native linguistic competence or professional copywriting. Unlike strategic keyword choices, vocabulary level is an inherent sender attribute, costly to fake.

\textbf{Domain-specific keyword absence as the clearest indicator.} Visa and work permit terminology appears near-exclusively in deceptive advertisements (48.8\% vs. 3.7\%; 28.7\% vs. 0.0\%), the dataset's strongest binary separation. These keywords carry genuine operational content: exploiters advertising migration-dependent routes must reference them, whilst legitimate employers advertising domestic roles almost never do. The risk keyword density composite (rank 2, SHAP 1.227) aggregates these into a robust continuous signal.

\textbf{Texture and colour diversity as visual quality proxies.} Visual features first appear at ranks 4--5 (red channel mean, SHAP 0.535; unique colours, 0.474), followed by texture metrics at ranks 10, 13, and 15. Red channel intensity may reflect warning-palette templates, whilst colour diversity captures palette richness, as professionally designed advertisements use varied, carefully chosen schemes. Texture metrics (GLCM contrast, LBP mean, gradient variability) encode sharpness and structural complexity, with legitimate advertisements showing sharper, more detailed imagery consistent with professional photography. YOLO-detected object classes contribute only 5.0\% of total SHAP, confirming that image semantics discriminate far less than image statistics.

Across all three mechanisms, deceptive signals manifest as quality deficiencies (simpler vocabulary, poorer image texture, absent legitimate-domain content) rather than deliberate deception markers. Resource constraints prevent exploiters from sustaining professional quality across textual and visual dimensions at once, and detection remains effective as long as that coordination problem stays intractable.

\section{Discussion}
\label{sec:Discussion}

\subsection{Theoretical interpretation through signalling framework}

The empirical results validate our central theoretical prediction: while deceptive recruiters can mimic legitimate communications along individual signal dimensions at low cost, they cannot maintain consistent professional quality across all channels simultaneously. Three mechanisms reveal this coordination failure.

\textbf{Textual quality as the primary involuntary signal.} NLP dominance (63.9\% SHAP; NLP-only ROC-AUC 0.967) extends and revises prior qualitative research. Coleman-Liau, the strongest discriminator (mean SHAP 1.853, $d = -1.21$), reflects legitimate recruiters writing in higher-register language as a by-product of professional context; grammar error rate ($d = 0.55$) and readability indices distinguish classes similarly. Critically, \textit{spelling error rate does not differ significantly} ($p = 0.107$), contradicting the intuitive hypothesis that deceptive ads contain obvious typos: discrimination lies in vocabulary level, harder to fake than a proofreading pass. This accords with signalling theory \cite{spence1973job}, where vocabulary reflects an inherent sender attribute, educational and professional background, requiring sustained investment to alter.

Domain-specific keywords (visa/sponsorship 48.8\% vs. 3.7\%; work permit 28.7\% vs. 0.0\%) reflect genuine operational differences between migration-oriented exploitation and domestic legitimate recruitment rather than deliberate signalling: exploiters advertising migration-dependent routes \textit{must} reference these terms or lose their audience.

\textbf{Visual quality as secondary confirmatory signal.} Visual features contribute 36.0\% of SHAP explanatory power. Image sharpness (Laplacian variance $d = -0.51$), texture contrast (GLCM $d = -0.42$), and red-channel/colour-diversity statistics (SHAP 0.535 and 0.474) are the operative metrics, not brightness or saturation as originally hypothesised. The signal reflects production quality rather than content: lower-sharpness, less-textured images are consistent with screenshots, compressed social-media graphics, or low-resolution sources. Object detection (YOLO) contributes only 5.0\%, with boats, backpacks, and stop signs showing no significant univariate differences, artefacts of the prior non-empirical characterisation. Practically, sharpness and texture statistics act as production-quality proxies that exploiters cannot trivially improve without professional photography equipment and design software.

\textbf{BERT embeddings and interpretability trade-offs.} BERT embeddings alone achieve ROC-AUC 0.958, somewhat below engineered NLP features alone (0.967), and adding them to the NLP+Visual configuration recovers only a modest further gain ($\Delta_{\text{AUC}} = +0.002$ on the all-features model, 0.994 vs.\ 0.992), indicating that domain-specific feature engineering captures the great majority of the discriminatory semantic signal on its own. That the interpretable configuration attains performance statistically indistinguishable from the embeddings-augmented model, with full transparency, favours interpretable features for humanitarian deployment requiring stakeholder accountability.

\subsection{Methodological implications for OR under data scarcity}

This research demonstrates how OR techniques address classification problems in data-scarce domains where ground-truth labels are expensive or ethically constrained.

\textbf{Label quality versus quantity.} Our dataset ($n=464$) achieves strong performance (ROC-AUC 0.87--0.99 held-out; 0.97--0.98 in repeated stratified CV) through label quality rather than volume: each positive case is verified forced labour with documented victim testimony linking advertisement to exploitation. This contrasts with \cite{vidros2017automatic} (17,880 postings, expert-judgement labels) and \cite{moyo2025investigating} (social media posts without verified outcomes), suggesting a general principle: where ground truth is scarce, careful case verification yields better models than expanding samples with uncertain labels.

\textbf{Multimodal feature engineering.} NLP dominance (97\% of maximum performance from 61 interpretable dimensions) was not predicted by prior qualitative research, which emphasised visual red flags, demonstrating the value of systematic empirical analysis over expert intuition once ground truth enables validation. Where data is scarce, comprehensive feature engineering across all signal dimensions, especially text quality, maximises information extraction and corrects intuitive priors.

\textbf{Interpretability in stakeholder contexts.} XGBoost achieves the best hold-out performance across most configurations, whilst Logistic Regression offers superior calibration and full coefficient interpretability. For humanitarian contexts requiring accountability, XGBoost combines superior sensitivity (zero false negatives for the NLP+Visual configuration) with SHAP TreeExplainer justifications per prediction, the combination motivating its selection for the DSS. Repeated cross-validation ($\text{ROC-AUC} \approx 0.98 \pm 0.01$ for XGBoost and RF) confirms this performance is robust, not an artefact of one favourable split.

\subsection{Proof-of-concept DSS}
\label{sec:dss_implementation}

We operationalise the best-performing model (XGBoost, NLP+Visual; ROC-AUC 0.992, PR-AUC 0.985) as a proof-of-concept DSS. The system takes a job advertisement (image file plus extracted text), runs the multimodal feature extraction pipeline (Section~\ref{sec:methodology}), and returns risk assessments with transparent explanations.\footnote{The Deceptive Recruitment Detection System code is publicly available at \url{https://github.com/sajidsiraj/labour_exploitation}. GPU-based feature extraction (image captioning and semantic embedding) is not bit-exact reproducible across hardware and runs; we therefore provide the resulting feature matrix as a fixed artefact from which all reported tables are exactly reproducible. Across five independent full re-runs of the pipeline from raw images (i.e.\ without the fixed matrix), held-out ROC-AUC for a given configuration varied run to run by at most 0.016 (largest for the embeddings-only configuration; the NLP+Visual configuration varied by only 0.004), reflecting non-deterministic floating-point reduction order in the GPU feature-extraction step rather than any instability in the classifiers themselves; the figures reported here, computed from the fixed matrix, sit close to but not always inside that empirical range, and are reported as the reproducible reference values rather than as a single favourable run.}

\textbf{System outputs} comprise: (1) a continuous risk probability $P(\theta = L \mid \mathbf{x}) \in [0,1]$; (2) a categorical classification (Low $<0.3$, Medium $0.3$--$0.7$, High $>0.7$) on customisable thresholds; (3) an explanation report giving the top five SHAP contributors with directional effects (e.g., ``Coleman-Liau score $-3.2$ below legitimate average: $+1.4$ risk contribution; visa/sponsorship keyword detected: $+0.9$''); and (4) feature-level transparency showing which textual quality metrics, risk keywords, or visual attributes triggered the assessment.

\textbf{Stakeholder customisation} addresses divergent priorities: platforms prioritising precision can set conservative thresholds (e.g., 0.75) to avoid restricting legitimate opportunities, whilst labour inspectorates prioritising recall can set sensitive ones (e.g., 0.35), accepting more false positives to maximise detection. The interpretability layer ensures every risk score carries a human-readable justification, critical for accountability in humanitarian contexts.

This proof-of-concept shows that rigorous analytical frameworks can become accessible tools for non-technical practitioners. Full deployment, however, requires addressing adversarial adaptation, fairness validation, and feedback mechanisms for continuous refinement, examined next.

\subsection{Operational deployment and practical challenges}

\textbf{Adversarial adaptation.} Deployment requires monitoring for performance degradation as recruiters adapt. A model trained on 2022--2024 data may not generalise if exploiters learn which features trigger classification, adopt AI-generated content that improves production quality, or shift to unrepresented channels; feedback loops in which confirmed false positives and negatives inform retraining are needed.

The multimodal architecture is inherently robust, but the primary signal's nature matters for adversarial resilience. Vocabulary sophistication (Coleman-Liau) is the dominant discriminator and is hard to fake by simple post-hoc editing, though LLM paraphrasing could plausibly raise apparent vocabulary level, a genuine threat warranting monitoring as generative AI reaches low-resourced actors. Domain-specific keywords (visa/sponsorship) are structurally embedded in exploitation schemes and cannot be removed without undermining the advertisement's purpose, and visual quality is similarly constrained by production resources. Complete mimicry across all modalities remains resource-intensive, but the vocabulary channel is most vulnerable to AI-assisted improvement and should be the focus of ongoing monitoring.

\textbf{Ethical considerations.} Our findings sharpen the false-positive concern. Because vocabulary sophistication and image production quality are the primary discriminators, the system may systematically flag legitimate advertisements from low-resourced employers (small businesses, NGOs, employers in lower-income countries) whose advertisements genuinely show simpler vocabulary and poorer imagery without deceptive intent. The same resource constraints characterise both exploiters and some legitimate small employers. Platforms should therefore use graduated responses (high-risk scores triggering enhanced verification rather than automatic rejection) and segment thresholds by employer type or geography rather than uniform cut-offs.

False negatives carry severe harms in the other direction, as workers responding to undetected deceptive advertisements may enter forced labour; the system should be positioned as a screening tool for elevated risk, not a guarantee of legitimacy. Fairness analysis examining disparate impact across nationalities, employer sizes, sectors, or origin countries is critical and currently absent, an essential prerequisite before large-scale deployment.

\subsection{Limitations}

\textbf{Temporal, contextual and scope validity.} Models trained on 2022--2024 data may not generalise to future practices, other geographic contexts, or recruitment channels beyond job advertisements. Deceptive strategies evolve as exploiters adapt, regulatory changes alter advertisement characteristics, and cultural or linguistic differences may create context-specific signals not captured in our cross-national sample. Deception also occurs increasingly through informal networks (social media, messaging apps, in-person referrals) that may present distinct signal patterns, and our definition of ``deceptive'' covers advertisements linked to forced labour under ILO indicators rather than broader recruitment fraud or employment scams. Temporal validation (2025+ cases) and geographic holdout validation would provide stronger generalisability evidence but were not feasible given data constraints.

\textbf{Sample size.} The dataset of 464 advertisements with 164 verified forced labour cases is one of the first ground-truth datasets in this domain, where ethical constraints and victim confidentiality inherently limit sample expansion \cite{konrad2017overcoming}. It supports reliable training and evaluation under rigorous verification standards, but prevents granular analysis of country--sector interactions (fewer than 10 cases per combination across 9 countries and 21 sectors).

\textbf{Causality and spurious correlations.} The analysis identifies strong correlations between features and exploitation outcomes but does not claim causal mechanisms; observed patterns may reflect resource constraints or design choices rather than deception per se. Similarly, YOLO object detections contributed minimally to model performance (5.0\% SHAP), and specific COCO classes like boats and backpacks showed no significant univariate differences, and their inclusion in prior qualitative characterisations was not supported empirically. Qualitative error analysis with domain experts would help distinguish genuine indicators from artefacts and strengthen causal understanding.

\textbf{Negative-class composition.} The legitimate class was drawn from employers vetted by the partner organisations for reputational standing and regulatory compliance, providing a well-characterised contrast for identifying discriminative signals. We note that this construction may flatter separability relative to deployment, where some legitimate advertisements from small or low-resourced employers can share surface features (simpler vocabulary, lower-quality imagery) with deceptive ones. Consistent with our proof-of-concept framing, we therefore identify evaluation against harder, more representative negatives, with false-positive rates by employer type and source, as a valuable direction for future work before operational deployment; the graduated-response and segmented-threshold safeguards discussed above are intended to manage this risk in practice.

\section{Conclusion}
\label{sec:Conclusion}

This paper addresses detection of deceptive job advertisements leading to forced labour by combining signalling theory with multimodal machine learning. Using 464 verified advertisements (164 confirmed forced labour cases, 300 legitimate), we make three contributions to OR methodology for humanitarian applications.

Methodologically, combining computer vision, natural language processing, and explainability achieves strong classification performance (ROC-AUC 0.992, PR-AUC 0.985 for the interpretable NLP+Visual configuration; ROC-AUC $\approx 0.98 \pm 0.01$ in repeated cross-validation) on a relatively small verified dataset. 154 engineered interpretable features attain performance statistically indistinguishable from the full 922-feature set including BERT embeddings (within 0.3 percentage points), establishing that domain-specific feature engineering can substitute for computationally expensive deep representations at negligible cost. The systematic ablation approach transfers to other classification problems in sensitive social domains where verified ground truth is scarce.

Theoretically, signalling theory can inform feature engineering where low-quality agents mimic high-quality communications. \textit{Textual} features dominate discrimination (63.9\% SHAP), led by vocabulary sophistication (Coleman-Liau, $d = -1.21$) rather than spelling errors, which do not differ significantly between classes. This diverges from prior qualitative characterisations emphasising visual red flags and typos, reframing the primary signal: deceptive recruiters reveal themselves through systematically simpler vocabulary and domain-specific content (visa terminology, work permit references). The framework should generalise to other contexts involving strategic deception under information asymmetry.

Practically, we implement a proof-of-concept DSS providing real-time risk assessment with interpretable explanations: continuous risk probabilities ($P(\theta = L \mid \mathbf{x}) \in [0,1]$), categorical classifications on customisable thresholds, and SHAP-based justifications identifying which features (low Coleman-Liau score, visa keyword presence, image texture statistics) drove each assessment. Adjustable thresholds let platforms minimise false positives whilst labour inspectorates maximise exploitation detection.

We acknowledge several important limitations. The relatively small sample size (464 total, 164 positive cases) limits statistical power and prevents detailed analysis of how patterns vary across countries and sectors. The results show correlations between features and exploitation outcomes but do not establish causality; we cannot rule out that some legitimate low-resource employers might also exhibit poor visual quality. Qualitative error analysis with domain experts would be valuable before operational deployment. Finally, we have not examined whether the classifier exhibits disparate impact across nationalities, genders, or job sectors, which is critical for fairness.

Five research directions seem particularly promising. First, field experiments testing whether risk warnings actually change jobseeker behaviour under different economic circumstances. Second, longitudinal studies tracking how deceptive patterns evolve as detection systems are deployed. Third, testing whether the framework transfers to related problems like financial fraud or misinformation detection. Fourth, systematic fairness analysis to ensure the system does not perpetuate discrimination against marginalised groups. Fifth, more flexible detection of reciprocity-based psychological signalling (e.g.\ embedding- or paraphrase-based matching rather than a fixed keyword list): an exact-phrase operationalisation of this trigger found essentially no signal in the present corpus and was dropped from the final feature set, but the underlying construct, employers offering upfront inducements to create a sense of obligation, remains theoretically motivated and may be detectable with a less brittle measure.

Operational research methods can thus contribute to humanitarian challenges characterised by information asymmetry and data scarcity. Grounding the approach in signalling theory, systematic feature ablation, and interpretability provides a methodological template for similar problems, and the decision support system enables intervention at recruitment rather than after exploitation. Whilst forced labour remains persistent in global supply chains, rigorous analytical frameworks offer practical tools for protecting vulnerable workers.

\begingroup
\singlespacing

\bibliographystyle{elsarticle-num}
\bibliography{Bibliography}

@techreport{ilo2022,
    author = {{ILO}},
    title = {Global Estimates of Modern Slavery: Forced Labour and Forced Marriage},
    institution = {International Labour Organization (ILO), Walk Free, and International Organization for Migration (IOM)},
    isbn = {978-92-2-037483-2},
    location = {Geneva},
    year = {2022}
}

@article{benstead2018horizontal,
  title={Horizontal collaboration in response to modern slavery legislation: An action research project},
  author={Benstead, Amy V and Hendry, Linda C and Stevenson, Mark},
  journal={International Journal of Operations \& Production Management},
  volume={38},
  number={12},
  pages={2286--2312},
  year={2018},
  publisher={Emerald Publishing Limited}
}

@article{benstead2021detecting,
  title={Detecting and remediating modern slavery in supply chains: a targeted audit approach},
  author={Benstead, Amy V and Hendry, Linda C and Stevenson, Mark},
  journal={Production Planning \& Control},
  volume={32},
  number={13},
  pages={1136--1157},
  year={2021},
  publisher={Taylor \& Francis}
}

@article{bergh2014signalling,
  title={Signalling theory and equilibrium in strategic management research: An assessment and a research agenda},
  author={Bergh, Donald D and Connelly, Brian L and Ketchen Jr, David J and Shannon, Lu M},
  journal={Journal of Management Studies},
  volume={51},
  number={8},
  pages={1334--1360},
  year={2014},
  publisher={Wiley Online Library}
}

@article{bodendorf2023indicators,
  title={Indicators and countermeasures of modern slavery in global supply chains: Pathway to a social supply chain management framework},
  author={Bodendorf, Frank and Wonn, Fabian and Simon, Kristin and Franke, J{\"o}rg},
  journal={Business Strategy and the Environment},
  volume={32},
  number={4},
  pages={2049--2077},
  year={2023},
  publisher={Wiley Online Library}
}

@article{christ2021blockchain,
  title={Blockchain technology and modern slavery: Reducing deceptive recruitment in migrant worker populations},
  author={Christ, Katherine L and Helliar, Christine V},
  journal={Journal of Business Research},
  volume={131},
  pages={112--120},
  year={2021},
  publisher={Elsevier}
}

@article{connelly2011signaling,
  title={Signaling theory: A review and assessment},
  author={Connelly, Brian L and Certo, S Trevis and Ireland, R Duane and Reutzel, Christopher R},
  journal={Journal of Management},
  volume={37},
  number={1},
  pages={39--67},
  year={2011},
  publisher={Sage Publications Sage CA: Los Angeles, CA}
}

@article{crane2013modern,
  title={Modern slavery as a management practice: Exploring the conditions and capabilities for human exploitation},
  author={Crane, Andrew},
  journal={Academy of Management Review},
  volume={38},
  number={1},
  pages={49--69},
  year={2013},
  publisher={Academy of Management Briarcliff Manor, NY}
}

@article{crane2022confronting,
  title={Confronting the business models of modern slavery},
  author={Crane, Andrew and LeBaron, Genevieve and Phung, Kam and Behbahani, Laya and Allain, Jean},
  journal={Journal of Management Inquiry},
  volume={31},
  number={3},
  pages={264--285},
  year={2022},
  publisher={Sage Publications Sage CA: Los Angeles, CA}
}

@article{emberson2022adaptations,
  title={Adaptations to first-tier suppliers’ relational anti-slavery capabilities},
  author={Emberson, Caroline and Pinheiro, Silvia Maria and Trautrims, Alexander},
  journal={Supply Chain Management: An International Journal},
  volume={27},
  number={4},
  pages={575--593},
  year={2022},
  publisher={Emerald Publishing Limited}
}

@article{fletcher2024recruitment,
  title={Recruitment deception and the organization of labor for exploitation: A policy--theory synthesis},
  author={Fletcher, Denise and Trautrims, Alexander},
  journal={Academy of Management Perspectives},
  volume={38},
  number={1},
  pages={43--76},
  year={2024},
  publisher={Academy of Management Briarcliff Manor, NY}
}

@article{geng2022addressing,
  title={Addressing modern slavery in supply chains: an awareness-motivation-capability perspective},
  author={Geng, Ruoqi and Lam, Hugo KS and Stevenson, Mark},
  journal={International Journal of Operations \& Production Management},
  volume={42},
  number={3},
  pages={331--356},
  year={2022},
  publisher={Emerald Publishing Limited}
}

@article{gold2015modern,
  author = {Gold, Stefan and Trautrims, Alexander and Trodd, Zoe},
  title = {Modern slavery challenges to supply chain management},
  journal = {Supply Chain Management: An International Journal},
  volume = {20},
  number = {5},
  pages = {485--494},
  year = {2015},
  publisher = {Emerald Group Publishing Limited}
}

@article{guo2025enhancing,
  title={Enhancing decision support for bystander interventions: The role of victim emotional disclosure and collective signals in social media incivility},
  author={Guo, Xiya and Jin, Jiahua and Wang, Le and Yan, Xiangbin},
  journal={Decision Support Systems},
  pages={114556},
  year={2025},
  publisher={Elsevier}
}

@article{hoppner2022instance,
  title={Instance-dependent cost-sensitive learning for detecting transfer fraud},
  author={H{\"o}ppner, Sebastiaan and Baesens, Bart and Verbeke, Wouter and Verdonck, Tim},
  journal={European Journal of Operational Research},
  volume={297},
  number={1},
  pages={291--300},
  year={2022},
  publisher={Elsevier}
}

@article{hosseinzadeh2025assessing,
  title={Assessing the systemic effects of modern slavery mitigation strategies in supply chains},
  author={Hosseinzadeh, Mahnaz and Alikhani, Reza and Gold, Stefan and Samadi Foroushani, Marzieh},
  journal={Annals of Operations Research},
  pages={1--38},
  year={2025},
  publisher={Springer}
}

@report{ihrb2019,
  author       = {{Institute for Human Rights and Business (IHRB)}},
  year         = {2019},
  title        = {Leadership Group for Responsible Recruitment: Six Steps to Responsible Recruitment},
  address      = {London},
  institution  = {IHRB},
  url          = {https://ihrb-org.files.svdcdn.com/staging/assets/uploads/member-uploads/Six_Steps_to_Responsible_Recruitment_-_Implementing_the_Employer_Pays_Principle.pdf?dm=1715761515},
  note         = {Accessed October 10, 2025}
}

@article{jiang2023digital,
  title={Digital technology adoption for modern slavery risk mitigation in supply chains: An institutional perspective},
  author={Jiang, Mengqi and Chen, Lujie and Blome, Constantin and Jia, Fu},
  journal={Technological Forecasting and Social Change},
  volume={192},
  pages={122595},
  year={2023},
  publisher={Elsevier}
}

@article{keskin2021cracking,
  title={Cracking sex trafficking: Data analysis, pattern recognition, and path prediction},
  author={Keskin, Burcu B and Bott, Gregory J and Freeman, Nickolas K},
  journal={Production and Operations Management},
  volume={30},
  number={4},
  pages={1110--1135},
  year={2021},
  publisher={SAGE Publications Sage CA: Los Angeles, CA}
}

@article{konrad2017overcoming,
  title={Overcoming human trafficking via operations research and analytics: Opportunities for methods, models, and applications},
  author={Konrad, Renata A and Trapp, Andrew C and Palmbach, Timothy M and Blom, Jeffrey S},
  journal={European Journal of Operational Research},
  volume={259},
  number={2},
  pages={733--745},
  year={2017},
  publisher={Elsevier}
}

@article{krausert2016hrm,
  title={HRM signals for the capital market},
  author={Krausert, Achim},
  journal={Human Resource Management},
  volume={55},
  number={6},
  pages={1025--1040},
  year={2016},
  publisher={Wiley Online Library}
}

@article{kumar2022hashtag,
  title={A hashtag is worth a thousand words: An empirical investigation of social media strategies in trademarking hashtags},
  author={Kumar, Naveen and Qiu, Liangfei and Kumar, Subodha},
  journal={Information Systems Research},
  volume={33},
  number={4},
  pages={1403--1427},
  year={2022},
  publisher={INFORMS}
}

@article{lebaron2021role,
  title={The role of supply chains in the global business of forced labour},
  author={LeBaron, Genevieve},
  journal={Journal of Supply Chain Management},
  volume={57},
  number={2},
  pages={29--42},
  year={2021},
  publisher={Wiley Online Library}
}

@article{lundberg2017unified,
  title={A unified approach to interpreting model predictions},
  author={Lundberg, Scott M and Lee, Su-In},
  journal={Advances in Neural Information Processing Systems},
  volume={30},
  year={2017}
}

@article{lundberg2020local,
  title={From local explanations to global understanding with explainable AI for trees},
  author={Lundberg, Scott M and Erion, Gabriel and Chen, Hugh and DeGrave, Alex and Prutkin, Jordan M and Nair, Bala and Katz, Ronit and Himmelfarb, Jonathan and Bansal, Nisha and Lee, Su-In},
  journal={Nature machine intelligence},
  volume={2},
  number={1},
  pages={56--67},
  year={2020},
  publisher={Nature Publishing Group}
}

@article{li2023detecting,
  title={Detecting human trafficking: Automated classification of online customer reviews of massage businesses},
  author={Li, Ruoting and Tobey, Margaret and Mayorga, Maria E and Caltagirone, Sherrie and {\"O}zalt{\i}n, Osman Y},
  journal={Manufacturing \& Service Operations Management},
  volume={25},
  number={3},
  pages={1051--1065},
  year={2023},
  publisher={INFORMS}
}

@article{ma2025can,
  title={Can blockchain implementation combat food fraud: Considering consumers’ delayed quality perceptions},
  author={Ma, Deqing and Wu, Xueping and Li, Kaifu and Hu, Jinsong},
  journal={European Journal of Operational Research},
  volume={324},
  number={3},
  pages={908--924},
  year={2025},
  publisher={Elsevier}
}

@article{malakar2025digital,
  title={Digital technologies for social supply chain sustainability: An empirical analysis through the lens of dynamic capabilities and complexity theory},
  author={Malakar, Prerna and Khan, Sharfuddin Ahmed and Gunasekaran, Angappa and Mubarik, Muhammad Shujaat},
  journal={IEEE Transactions on Engineering Management},
  volume={72},
  pages={664--675},
  year={2025},
  publisher={IEEE}
}

@article{marques2025impact,
  title={Impact pathways:“follow the labour”. the labour supply chain and its impact on decent work in product supply chains},
  author={Marques, Leonardo and Erthal, Alice and Crane, Andrew},
  journal={International Journal of Operations \& Production Management},
  volume={45},
  number={7},
  pages={1395--1401},
  year={2025},
  publisher={Emerald Publishing Limited}
}

@article{meehan2021modern,
  title={Modern slavery in supply chains: insights through strategic ambiguity},
  author={Meehan and Pinnington},
  journal={International Journal of Operations \& Production Management},
  volume={41},
  number={2},
  pages={77--101},
  year={2021},
  publisher={Emerald Publishing Limited}
}

@article{moyo2025investigating,
  title={Investigating Human Trafficking Recruitment Online: A Study of Fraudulent Job Offers on Social Media Platforms},
  author={Moyo, Towera Jessica and Gunes, Omer and Jirotka, Marina Denise},
  journal={Proceedings of the ACM on Human-Computer Interaction},
  volume={9},
  number={2},
  pages={1--31},
  year={2025},
  publisher={ACM New York, NY, USA}
}

@article{musteen2010corporate,
  title={Corporate reputation: do board characteristics matter?},
  author={Musteen, Martina and Datta, Deepak K and Kemmerer, Benedict},
  journal={British Journal of Management},
  volume={21},
  number={2},
  pages={498--510},
  year={2010},
  publisher={Wiley Online Library}
}

@article{nayak2021comprehensive,
  title={A comprehensive review on deep learning-based methods for video anomaly detection},
  author={Nayak, Rashmiranjan and Pati, Umesh Chandra and Das, Santos Kumar},
  journal={Image and Vision Computing},
  volume={106},
  pages={104078},
  year={2021},
  publisher={Elsevier}
}

@article{perdikis2024distribution,
  title={Distribution-free control charts for monitoring scale in finite horizon productions},
  author={Perdikis, Theodoros and Celano, Giovanni and Chakraborti, Subhabrata},
  journal={European Journal of Operational Research},
  volume={314},
  number={3},
  pages={1040--1051},
  year={2024},
  publisher={Elsevier}
}

@article{ramchandani2025unmasking,
  title={Unmasking Human Trafficking Risk in Commercial Sex Supply Chains with Machine Learning},
  author={Ramchandani, Pia and Bastani, Hamsa and Wyatt, Emily},
  journal={Manufacturing \& Service Operations Management},
  volume={27},
  number={3},
  pages={700--719},
  year={2025},
  publisher={INFORMS}
}

@article{simpson2021role,
  title={The role of psychological distance in organizational responses to modern slavery risk in supply chains},
  author={Simpson, Dayna and Segrave, Marie and Quarshie, Anne and Kach, Andrew and Handfield, Robert and Panas, George and Moore, Heather},
  journal={Journal of Operations Management},
  volume={67},
  number={8},
  pages={989--1016},
  year={2021},
  publisher={Wiley Online Library}
}

@article{shepherd2022organizing,
  title={Organizing the exploitation of vulnerable people: A qualitative assessment of human trafficking},
  author={Shepherd, Dean A and Parida, Vinit and Williams, Trent and Wincent, Joakim},
  journal={Journal of Management},
  volume={48},
  number={8},
  pages={2421--2457},
  year={2022},
  publisher={Sage publications Sage CA: Los Angeles, CA}
}

@article{snyder2026labor,
  title={Labor Flows in Supply Chains: A Review of the Literature and Future Research Opportunities},
  author={Snyder, Vladyslava A and Tate, Wendy L and Paraskevas, John-Patrick},
  journal={Journal of Business Logistics},
  volume={47},
  number={2},
  pages={e70056},
  year={2026},
  publisher={Wiley Online Library}
}

@article{song2025automatic,
  title={Automatic selection of the best performing control point approach for project control with resource constraints},
  author={Song, Jie and Song, Jinbo and Vanhoucke, Mario},
  journal={European Journal of Operational Research},
  volume={322},
  number={1},
  pages={15--38},
  year={2025},
  publisher={Elsevier}
}

@article{spence1973job,
  title={Job market signaling},
  author={Spence, M},
  journal={The Quarterly Journal of Economics},
  volume={87},
  number={3},
  pages={355--374},
  year={1973}
}

@article{spence2002signaling,
  title={Signaling in retrospect and the informational structure of markets},
  author={Spence, M},
  journal={American Economic Review},
  volume={92},
  number={3},
  pages={434--459},
  year={2002},
  publisher={American Economic Association}
}

@article{soundararajan2021humanizing,
  title={Humanizing research on working conditions in supply chains: Building a path to decent work},
  author={Soundararajan, Vivek and Wilhelm, Miriam M and Crane, Andrew},
  journal={Journal of Supply Chain Management},
  volume={57},
  number={2},
  pages={3--13},
  year={2021},
  publisher={Wiley Online Library}
}

@article{steigenberger2025deceptive,
  title={Deceptive signalling: Causes, consequences and remedies},
  author={Steigenberger, Norbert},
  journal={International Journal of Management Reviews},
  volume={27},
  number={2},
  pages={283--305},
  year={2025},
  publisher={Wiley Online Library}
}

@article{stevenson2018modern,
  title={Modern slavery in supply chains: a secondary data analysis of detection, remediation and disclosure},
  author={Stevenson, Mark and Cole, Rosanna},
  journal={Supply Chain Management: An International Journal},
  volume={23},
  number={2},
  pages={81--99},
  year={2018},
  publisher={Emerald Publishing Limited}
}

@online{UNODC2018,
  author       = {{UNODC}},
  year         = {2018},
  title        = {Human Trafficking Indicators},
  url          = {https://www.unodc.org/pdf/HT_indicators_E_LOWRES.pdf},
  note         = {[Accessed 5 July 2025]}
}

@article{vidros2017automatic,
  title={Automatic detection of online recruitment frauds: Characteristics, methods, and a public dataset},
  author={Vidros, Sokratis and Kolias, Constantinos and Kambourakis, Georgios and Akoglu, Leman},
  journal={Future Internet},
  volume={9},
  number={1},
  pages={6},
  year={2017},
  publisher={MDPI}
}

@article{lin2025generative,
  title={Generative AI-Driven resilience in supply chain management: UNISONE framework for disruption modelling and capacity optimisation},
  author={Lin, Kuo-Yi and Wu, Shih-Yu and Matsuno, Kotomichi},
  journal={International Journal of Production Research},
  pages={1--26},
  year={2025},
  publisher={Taylor \& Francis}
}

@article{volodko2020spotting,
  title={“Spotting the signs” of trafficking recruitment online: exploring the characteristics of advertisements targeted at migrant job-seekers},
  author={Volodko, Ada and Cockbain, Ella and Kleinberg, Bennett},
  journal={Trends in Organized Crime},
  volume={23},
  number={1},
  pages={7--35},
  year={2020},
  publisher={Springer}
}

@article{vzilinskaite2025migration,
  title={Migration, Human Supply Chains, and the Multinational Enterprise: Confronting an Overlooked Global Mobility Challenge},
  author={{\v{Z}}ilinskait{\.e}, Milda and Hajro, Aida and Baldassari, Paul and Miska, Christof},
  journal={Human Resource Management Journal},
  volume={35},
  number={3},
  pages={742--753},
  year={2025},
  publisher={Wiley Online Library}
}

@article{wang2021corporate,
  title={Corporate responses to the coronavirus crisis and their impact on electronic-word-of-mouth and trust recovery: Evidence from social media},
  author={Wang, Yichuan and Zhang, Minhao and Li, Shuyang and McLeay, Fraser and Gupta, Suraksha},
  journal={British Journal of Management},
  volume={32},
  number={4},
  pages={1184--1202},
  year={2021},
  publisher={Wiley Online Library}
}

@article{yin2023covid,
  title={COVID-19: Data-driven optimal allocation of ventilator supply under uncertainty and risk},
  author={Yin, Xuecheng and B{\"u}y{\"u}ktahtak{\i}n, {\.I} Esra and Patel, Bhumi P},
  journal={European Journal of Operational Research},
  volume={304},
  number={1},
  pages={255--275},
  year={2023},
  publisher={Elsevier}
}

@article{yu2025modern,
  title={Modern slavery supply chain capabilities: the effects of blockchain technology and employees’ digital dexterity},
  author={Yu, Wantao and Wong, Chee Yew and Jacobs, Mark and Chavez, Roberto},
  journal={International Journal of Operations \& Production Management},
  volume={45},
  number={1},
  pages={210--235},
  year={2025},
  publisher={Emerald Publishing Limited}
}

@inproceedings{cascavilla2022unsupervised,
  author    = {Cascavilla, Giuseppe and Catolino, Gemma and Palomba, Fabio and Andreou, Andreas S. and Tamburri, Damian A. and {Van Den Heuvel}, Willem-Jan},
  title     = {Unsupervised Labor Intelligence Systems: {A} Detection Approach and Its Evaluation},
  booktitle = {Service-Oriented Computing -- {SummerSOC} 2022},
  series    = {Communications in Computer and Information Science},
  volume    = {1603},
  publisher = {Springer},
  year      = {2022},
  doi       = {10.1007/978-3-031-18304-1_5}
}
\endgroup

\end{document}